\documentclass[%
 reprint,
superscriptaddress,
 amsmath,amssymb,
 aps,
]{revtex4-2}

\usepackage{graphicx}
\usepackage{dcolumn}
\usepackage{bm}
\usepackage{hyperref}

\usepackage{array}
\usepackage{tabularx}

\begin{document}
\preprint{APS/123-QED}

\title{Electron and Hole Mobility of SnO$_2$ from Full-Band Electron--Phonon and Ionized Impurity Scattering~Computations}

\author{Zhen Li}
\email{Zhen.Li.2@warwick.ac.uk}
 \affiliation{School of Engineering, University of Warwick, Coventry, CV4 7AL, UK.}
 
\author{Patrizio Graziosi}%
 \affiliation{Institute of Nanostructured Materials, CNR, Bologna, Italy.}
 
\author{Neophytos Neophytou}
 \affiliation{School of Engineering, University of Warwick, Coventry, CV4 7AL, UK.}

\date{\today}

\begin{abstract}

Mobility is a key parameter for SnO$_2$, which is extensively studied as a practical transparent oxide $n$-type semiconductor. In experiments, the mobility of electrons in bulk SnO$_2$ single crystals varies from 70 to 260 cm$^2$V$^{-1}$s$^{-1}$ at room temperature. Here, we calculate the mobility as limited by electron--phonon and ionized impurity scattering by coupling the Boltzmann transport equation with density functional theory electronic structures. The linearized Boltzmann transport equation is solved numerically beyond the commonly employed constant relaxation-time approximation by taking into account all energy and momentum dependencies of the scattering rates. Acoustic deformation potential and polar optical phonons are considered for electron--phonon scattering, where polar optical phonon scattering is found to be the main factor which determines the mobility of both electrons and holes at room temperature. The calculated phonon-limited electron mobility is found to be 265 cm$^2$V$^{-1}$s$^{-1}$, whereas that of holes is found to be 7.6 cm$^2$V$^{-1}$s$^{-1}$. We present the mobility as a function of the carrier concentration, which shows the upper mobility limit. The large difference between the mobilities of $n$-type and $p$-type SnO$_2$ is a result of the different effective masses between electrons and holes.

\end{abstract}

\maketitle


\section{Introduction}

Tin oxide (SnO$_2$) is a critically important $n$-type semiconductor with a relatively high mobility and a wide band gap ($E_{\rm g}$ = 3.6--3.7 eV) \cite{Sundaram1981,Melsheimer1985}. Due to its good electrical, optical, and electrochemical properties, SnO$_2$ has been extensively exploited in various state-of-the-art applications: perovskite solar cells \cite{Xiong2018}, as both compact layers and mesoporous layers for transparent electrodes; lithium-ion batteries \cite{Chen2013}, as promising candidates to serve as the anode material due to their high theoretical capacity; gas sensors \cite{Das2014}, as the most commonly used commercial material \cite{Staerz2020}; photocatalytic applications \cite{Sun2022}, as photocatalysts in organic pollutant degradation, water splitting, Cr(VI) reduction, CO$_2$ reduction, air purification, and photocatalytic sterilization; thermoelectric materials \cite{Chen2011}, as ceramic thermocouples to replace noble-metal thermocouples that are unable to withstand the harsh environments inside the hot sections of turbine engines used for power generation and propulsion. 

Mobility is a key factor in charge transport since it describes how the motion of an electron is affected by an applied electric field. As such, it is an important element in the design of modern devices. In experiments, the electron mobilities of bulk SnO$_2$ single crystals vary from 70 to 260 cm$^2$V$^{-1}$s$^{-1}$ at room temperature \cite{Morgan1966,Fonstad1971,Galazka2014}, while SnO$_2$ thin films show lower electron mobilities from 25 to 130 cm$^2$V$^{-1}$s$^{-1}$ \cite{Okude2008,Toyosaki2008,White2009,Mun2015,Fukumoto2020}. The large variation is a result of the many carrier scattering processes that take place beyond the intrinsic electron--phonon, such as scattering by ionized impurities \cite{Chattopadhyay1981}, neutral impurities \cite{Erginsoy1950}, grain boundaries \cite{Blatter1986}, and dislocations \cite{Jaszek2001}. To  properly evaluate the intrinsic mobility of the material, as well as that of the doped material, we need to calculate its electronic transport using full band electronic structure details, but also consider scattering processes that include the entire energy and momentum dependence of the scattering rates, beyond the constant relaxation time approximation. The latter is one of the earliest and most common approaches \cite{Li2018,Madsen2018,Li2019}, especially in the context of high-throughput computational searches targeting electronic transport properties \cite{Gorai2017}. However, it introduces an arbitrary uncertainty upon the choice of the scattering time.\cite{Graziosi2019}

Conventional mobility models suppress atomic-scale detail, relying on deformation potentials and either effective-mass theory or bulk energy bands to describe electron velocities \cite{Li2021,Graziosi2022}. They do not necessarily represent the behavior of a practical device where more significant extrinsic scattering mechanisms are present. Despite this, the intrinsic mobility of a material still provides an upper limit to the material's performance. Previous calculations reported a lower electron mobility (187 cm$^2$V$^{-1}$s$^{-1}$) compared to the highest reported experimental values, as well as hole mobility (14.1 cm$^2$V$^{-1}$s$^{-1}$) considering only electron--phonon scattering using density functional theory calculations \cite{Hu2019}. Other works include ionized impurity scattering using the empirical Brooks–Herring–Dingle formula, but use a fixed value (260 cm$^2$V$^{-1}$s$^{-1}$) for the phonon-limited mobility of $n$-type SnO$_2$~\cite{Fukumoto2020}. For proper mobility evaluation of the doped material, it is necessary to use a full-band numerical approach to compute the intrinsic mobility for both electron--phonon and ionized impurity scattering.

\begin{figure*}
\includegraphics[width=11.5 cm]{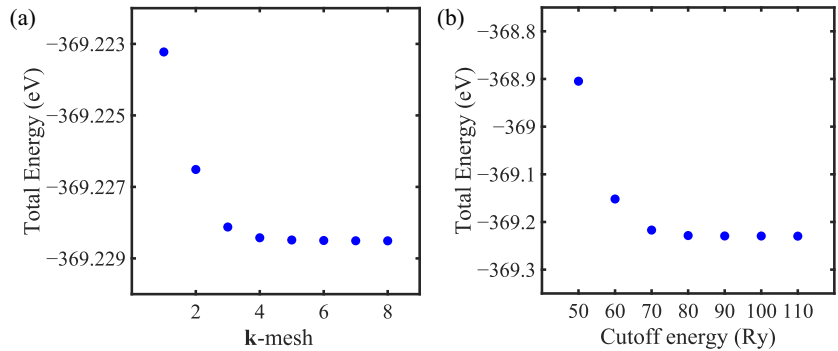}
\caption{(\textbf{a}) Total energy versus {\bf k} meshes along the x and y directions, where the {\bf k} mesh along the z direction is set to 10. (\textbf{b}) Total energy versus cutoff energy. \label{fig1}}
\end{figure*} 

In this work, we use the full energy and momentum dependencies of electron--phonon and ionized impurity scattering to compute the mobility for both electrons and holes in SnO$_2$. We first compute the band structures from density functional theory, from which we also extract the density of states' effective mass and the conductivity effective mass. Then, we calculate the acoustic deformation potential and polar optical phonon scattering rates. Finally, we use those rates within the linearized Boltzmann transport equation, which is solved numerically beyond the constant relaxation-time approximation to calculate the mobility of SnO$_2$ as a function of the carrier concentration.

\section{Computational Methods}

The electronic band structure is calculated from density functional theory (DFT) using the Quantum ESPRESSO package \cite{Giannozzi2009}. The optimized norm-conserving Vanderbilt (ONCV) pseudopotentials are used for Sn and O under the generalized gradient approximation (GGA) with the Perdew--Burke--Ernzerhof (PBE) functional \cite{Perdew1996,Hamann2013}. The 6~$\times$~6~$\times$~10 and 120~$\times$~120~$\times$~200 Monkhorst-Pack $\bf k$ meshes are used for structure relaxation and electronic band structure calculations, respectively. The cutoff energy of plane waves is set to 80~Ry. All of the parameters have been tested to be sufficient in obtaining converged results with the differences between the total energy being less than 0.001 eV/atom, as shown in Figure~\ref{fig1}.

For electronic transport calculations and relevant quantities including the scattering rates, mobility, transport distribution function, band velocity, density of states, and carrier concentrations, we use the ElecTra code \cite{Graziosi2019,Graziosi20222}, our home-developed, open-source code which solves the linearized Boltzmann transport equation in the relaxation time approximation for charge carriers in a full-band electronic structure of arbitrary complexity, including their energy, momentum, and band-index dependence.

\section{Results and Discussions}

\begin{figure*}
\includegraphics[width=10.5 cm]{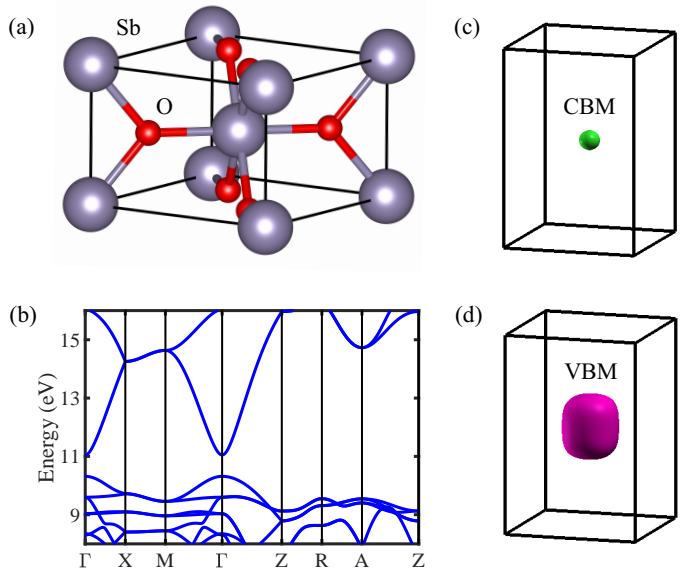}
\caption{(\textbf{a}) Lattice structure for SnO$_2$. (\textbf{b}) Band structure for SnO$_2$ along high-symmetry lines. (\textbf{c})~Fermi surface at energy $E$ = 0.2 eV above the conduction band minimum (CBM). (\textbf{d}) Fermi surface at energy $E$ = 0.2 eV below the valance band maximum (VBM).}
\label{fig2}
\end{figure*}

SnO$_2$ is a Rutile structure and crystallizes in the tetragonal P4$_2$/mnm space group, as shown in Figure \ref{fig2}a. The calculated lattice parameters are $a = b$ = 4.81 \AA, $c$ = 3.23 \AA, indicating a slight 1.5\% overestimation with respect to the available experimental value of $a = b$ = 4.74 \AA, $c$ = 3.19 \AA \cite{Madelung2004}, which is the general tendency of GGA \cite{Stampfl2001}. Both the valance band maximum (VBM) and conduction band minimum (CBM) are located at the $\Gamma$ point, as shown in Figure~\ref{fig2}b. The band gap $E_{\rm g}$ is calculated to be 0.734 eV, lower than experimental values ($E_{\rm g}$ = 3.6--3.7 eV) \cite{Sundaram1981,Melsheimer1985}, but in good agreement with previous calculations (\mbox{$E_{\rm g}$ = 0.832 eV}) using GGA \cite{El2013}. There is a known problem with the underestimation of the band gap using the GGA pseudopotentials \cite{Perdew2017}. This shortage can, in principle, be overcome by using Heyd--Scuseria--Ernzerhof (HSE) hybrid functionals \cite{Heyd2003}, $GW$ method \cite{Aryasetiawan1998}, \mbox{GGA + U} method \cite{Anisimov1991}, meta-GGA functionals \cite{Isaacs2018}, or the Tran–Blaha-modified Becke–Johnson (TB-mBJ) exchange potential approximations \cite{Tran2009}. We note, however, that since the band gap of this material is large enough, bipolar transport is suppressed, and we have considered the conduction bands and valence bands separately in the transport calculations. With regards to the accuracy of the band structure parameters, previous work has compared the band structure using GGA and TB-mBJ corrections for SnO$_2$ \cite{El2013} and found very similar overall behavior of the band structures, except for the value of band gap. Thus, we take that GGA is reliable enough to describe the overall behavior of the band structures to be used in our transport calculations. 

\subsection{Effective Mass Extraction Method}

The flat valance band and dispersive conduction band indicate heavy hole but light electron states, as shown in Figures 2(c) and 2(d), respectively. Here, we use our home-developed Effective Mass Finder (EMAF) code to calculate the two relevant effective masses for electrons and holes as described in references \cite{Neophytou2010,Graziosi2020}: the density of states effective mass ($m_{\rm DOS}$) and the conductivity effective mass $m_{\rm cond}$. We compute the $m_{\rm DOS}$ as the effective mass of an isotropic parabolic band that gives the same carrier density as the actual band structure. We evaluate $m_{\rm cond}$ as the effective mass of an isotropic parabolic band, which maps the average velocity of the band states weighted by their contribution to transport. For this, we employ a simple ballistic field effect transistor model, extract the average injection velocity in the sub-threshold regime, and map that velocity to a parabolic band, which provides the same injection velocity.

\begin{figure*}
\includegraphics[width=11.2 cm]{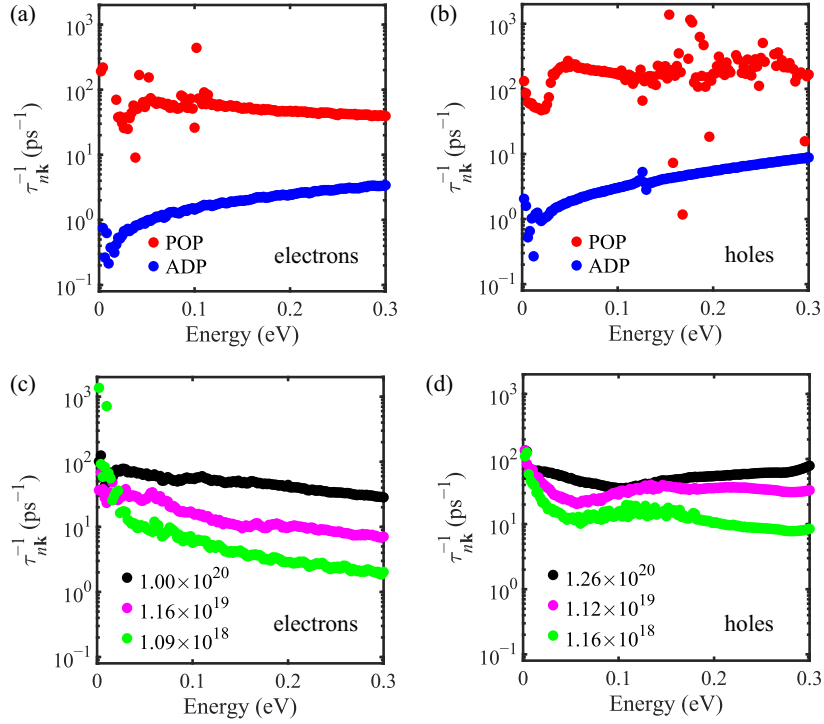}
\caption{Scattering rates arising from polar optical phonon (POP) and acoustic deformation potential (ADP) scattering processes in SnO$_2$ for (\textbf{a}) electrons and (\textbf{b}) holes at 300 K. Ionized impurity scattering rates in SnO$_2$ for (\textbf{c}) electrons and (\textbf{d}) holes at different impurity concentrations. The conduction band minimum is set to zero eV in (\textbf{a},\textbf{c}), while the valence band maximum is set to zero eV in (\textbf{b},\textbf{d}).
\label{fig3}}
\end{figure*}  

In detail, the process is as follows (using the conduction band as an example): We consider the non-degenerate regime, in which the carrier concentration $n$ can be expressed~as: 
\begin{equation}
n=N_{\rm C} e^ {\frac{E_{\rm F}-E_0}{k_{\rm B} T}}
\end{equation}
where $E_{\rm F}$ is the Fermi level, $E_0$ is the energy of the band edge, $k_{\rm B}$ is the Boltzmann constant, $T$ is the temperature, and $N_{\rm C}$ is the effective density of states calculated as:
\begin{equation}
N_{\rm C}=2(\frac{m_{\rm DOS} k_{\rm B} T}{2\pi \hbar ^2})^{\frac{3}{2}} 
\end{equation}

For a generic numerical band structure, the carrier concentration $n$ can be calculated~as:
\begin{equation}
n=\frac{2}{(2\pi)^3} \sum _{{\rm \bf k},n}f_{E_{({\rm \bf k},n)}} ) dV_{\rm \bf k}
\end{equation}
where the sum runs over all the {\bf k} points and bands in the first Brillouin zone, $f_{E_{({\rm \bf k},n)}}$ is the Fermi--Dirac distribution, and $dV_{\rm \bf k}$ is the volume element in {\bf k} space, which usually depends only on the mesh.

We place the Fermi level in non-degenerate conditions (in the band gap) and compute the carrier concentration $n$ using Equation (3). We then match $n$ with that calculated by using the non-degenerate statistics and the effective density of states $N_{\rm C}$ in Equations (1) and (2). A value for a parabolic mass which provides the same charge density as the whole band structure can then be extracted from $N_{\rm C}$. The so-calculated $m_{\rm DOS,e}$ = 0.2305 $m_0$ and $m_{\rm DOS,h}$ = 1.7268 $m_0$.

For the conductivity effective mass, we essentially calculate the average uni-directional velocity of all the 3D band structure states, where the contribution of each state is averaged by its occupancy under non-degenerate conditions. We then map that velocity value to the velocity from a parabolic band. The parabolic effective mass that is needed for this mapping is the conductivity mass we are looking for. The parabolic mass extracted in this way will collectively incorporate all details of the band structure related to transport. The conductivity effective mass $m_{\rm cond}$ is calculated from the injection velocity $v_{\rm inj}$ as \cite{Neophytou2010}:
\begin{equation}
m_{\rm cond}=\frac{2k_{\rm B} T}{\pi v_{\rm inj}^2}
\end{equation}

We then assume that the material of interest is the channel of a ballistic field effect transistor (FET). The injection velocity $v_{\rm inj}$, which depends only on the band structure and the temperature, is extracted by dividing the subthreshold current of a ballistic FET by the charge density occupation as \cite{Rahman2003}:

\begin{equation}
v_{\rm inj}=\frac{I_{\rm FET}} {\frac{e}{2\pi^3}\sum_{{\rm \bf k},n} f_{(E_{{\rm \bf k},n} -E_{\rm F,S})} dV_{\rm \bf k}}
\end{equation}
where $E_{\rm F,S}$ is the Source Fermi level, and $I_{\rm FET}$ is the FET current density. 

Here, we assume injection of carriers from the source contact of the FET and a high drain voltage, such that the Fermi level in the drain is much lower compared to that in the source. In this case, the injection of carriers from the drain is negligible and can be omitted, and thus the total FET current is just the source current. This current can be simply computed by counting all positive velocity going states weighted by their occupancy:
\begin{equation}
I_{\rm FET}=\frac{e}{2\pi^3} \sum_{{\rm \bf k},n} f_{(E_{{\rm \bf k},n} -E_{\rm F,S})} |v_{{\rm \bf k},n}| dV_{\rm \bf k}
 \end{equation}
where $|v_{{\rm \bf k},n}|$ is the band velocity in absolute terms to account only for positive traveling states. We perform this calculation in the three major orientations, $x$, $y$, and $z$, and then average (inversely) the three masses to obtain an overall conductivity mass as: 
\begin{equation}
m_{\rm cond}=\frac{3}{\frac{1}{m_{{\rm cond},x}}+\frac{1}{m_{{\rm cond},y}}+\frac{1}{m_{{\rm cond},z}}}
\end{equation}
In this way, the calculated $m_{\rm cond,e}$ = 0.2011 $m_0$ and $m_{\rm cond,h}$ = 1.7659 $m_0$.

\subsection{Scattering Rates}

The electron--phonon scattering rates, due to the acoustic deformation potential, can be computed by evaluating the transition rates  $|S_{\bf{k,k'}}^{\rm{ADP}}|$ between the initial $\bf{k}$ and final $\bf{k'}$ states and can be extracted using Fermi's golden rule as \cite{Lundstrom2000,Neophytou2020,Li2021}:
\begin{equation}
|S_{\bf{k,k'}}^{\rm{ADP}}|=\frac{\pi}{\hbar} D_{\rm{ADP}}^2 \frac{k_B T}{\rho v_{\rm s}^2} g(E)
\end{equation}
Here, $D_{\rm{ADP}}$ is the acoustic deformation potential, where 8.17 eV and 2.06 eV are used for electrons and holes in SnO$_2$ \cite{Hu2019}, respectively. $\rho$ is the mass density. $v_{\rm s}$ is the sound velocity of the material, where 4.3 km/s is used \cite{Turkes1980}. $g(E)$ is the density of states for the initial state.

The polar optical phonon scattering rates due to the Fr\"{o}hlich interaction, $|S_{\bf{k,k'}}^{\rm{POP}}|$, can be computed from the dielectric constants which capture the matrix element in a polarizable continuum as \cite{Lundstrom2000,Neophytou2020}:
\begin{equation}
|S_{{\bf k},{\bf k'}}^{\rm POP}| = \frac{\pi q_0^2 \omega }{|{\bf k}-{\bf k'}|^2 \varepsilon _0} (\frac{1}{k_\infty}-\frac{1}{k_0}) (N_{\omega} + \frac{1}{2} \mp \frac{1}{2}) g(E \pm \hbar \omega)
\end{equation}
Here, $\omega $ is the dominant frequency of polar optical phonons over the whole Brillouin zone, which has been validated to be a satisfactory approximation \cite{Samsonidze2018}, where 27.3 meV is used \cite{Hu2019}. $\varepsilon_0$ is the vacuum dielectric constant. $k_0$ and $k_\infty$ are the static and high-frequency dielectric constants, respectively, where $k_0$ = 11.6 and $k_\infty$ = 4.1 are used \cite{Hu2019}. $N_{\omega}$ is the Bose--Einstein phonon statistical distribution. 

Figure~\ref{fig3}a,b show the calculated scattering rates for acoustic deformation potential (ADP) scattering and polar optical phonon (POP) scattering for electrons and holes, respectively. The POP scattering rates for both electrons and holes are much larger than the ADP scattering rates, as expected from a polar material. On the other hand, compared to electrons, holes have larger ADP and POP scattering rates. This can be understood from the density of states for electrons and holes:
\begin{equation}
g(E) = \frac{\sqrt{2{m_{\rm DOS}^3}E}}{\pi ^2{\hbar}^3}
\end{equation}
Compared to electrons ($m_{\rm DOS,e}$ = 0.2305 $m_0$), holes have a much larger density of states effective masses ($m_{\rm DOS,h}$ = 1.7268 $m_0$) and much larger density of states, which results in the larger ADP and POP scattering rates.

In addition to electron--phonon scattering, the Brooks--Herring model is used to describe the elastic scattering rates due to ionized dopants \cite{Jacoboni1983}. The scattering rates due to ionized impurity scattering (IIS) are given by \cite{Nag1980}:
\begin{eqnarray}
|S_{{\bf k},{\bf k'}}^{\rm IIS}|  = \frac{2\pi}{\hbar} \frac{Z^2 e^4}{k_s^2 \varepsilon _0^2} \frac{N_{\rm imp}}{(|{\bf k}- {\bf k'}|^2+ L_{\rm D}^{-2})^2} g(E)
\end{eqnarray}
where $Z$ is the electric charge of the ionized impurity, $N_{\rm imp}$ is the density of the ionized impurities, and $L_{\rm D}$ is the Debye screening length defined as \cite{Nag1980}:
\begin{eqnarray}
L_{\rm D}=\sqrt{\frac{k_s \varepsilon_0}{e} (\frac{\partial E_{\rm F}}{\partial n})}
\end{eqnarray}
where $n$ is the carrier concentration. The partial derivative is computed at any Fermi level and temperature as follows: the carrier concentration is computed by evaluating the integral over the energy of the product of DOS and Fermi distribution, then the Fermi level is moved by 1 meV, and the carrier concentration is computed again. From this finite $E_{\rm F}$ difference, the partial derivative is obtained.

\begin{figure*}
\includegraphics[width=11.2 cm]{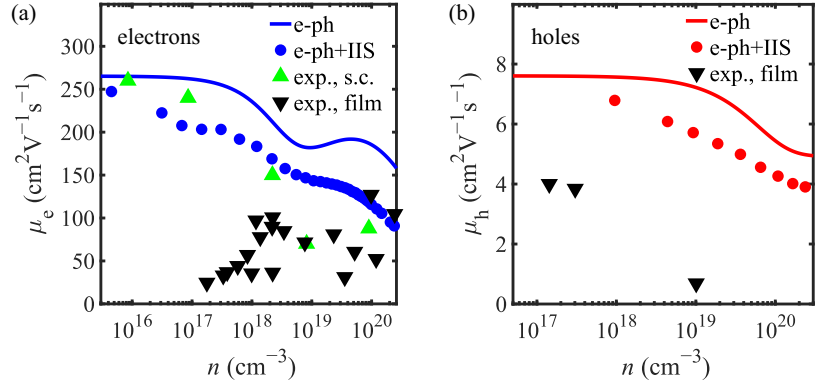}
\caption{The calculated mobility versus carrier concentration for (\textbf{a}) electrons and (\textbf{b}) holes in SnO$_2$. Phonon-limited (solid lines) and phonon plus ionized impurity scattering (dotted lines) are shown. Experimental measurements from single crystals (s.c., green triangles) and thin films (black triangles) are also indicated. References for the data in (\textbf{a}): single crystals (refs.~\protect\cite{Morgan1966,Fonstad1971}) and thin films (\mbox{refs. \protect\cite{Okude2008,Toyosaki2008,White2009,Mun2015,Fukumoto2020}}) for electrons. References for the data in (\textbf{b}): films (refs. \protect\cite{Yu2013,Fu2021}) for holes.
\label{fig4}}
\end{figure*} 

Figure~\ref{fig3}c,d show the calculated ionized impurity scattering rates at different impurity concentrations for electrons and holes in SnO$_2$, respectively. Compared to the electron--phonon scattering rates, the ionized impurity scattering rates for electrons at high impurity concentrations, e.g., at 10$^{20}$ cm$^{-3}$, are comparable to the POP scattering rates. However, for holes, even at a high impurity concentration ($1.26 \times 10^{20}$ cm$^{-3}$), the ionized impurity scattering rates are still lower than the POP scattering rates. Thus, the POP will always dominate the scattering for holes in SnO$_2$ at room temperature.

\begin{figure*}
\includegraphics[width=11.2 cm]{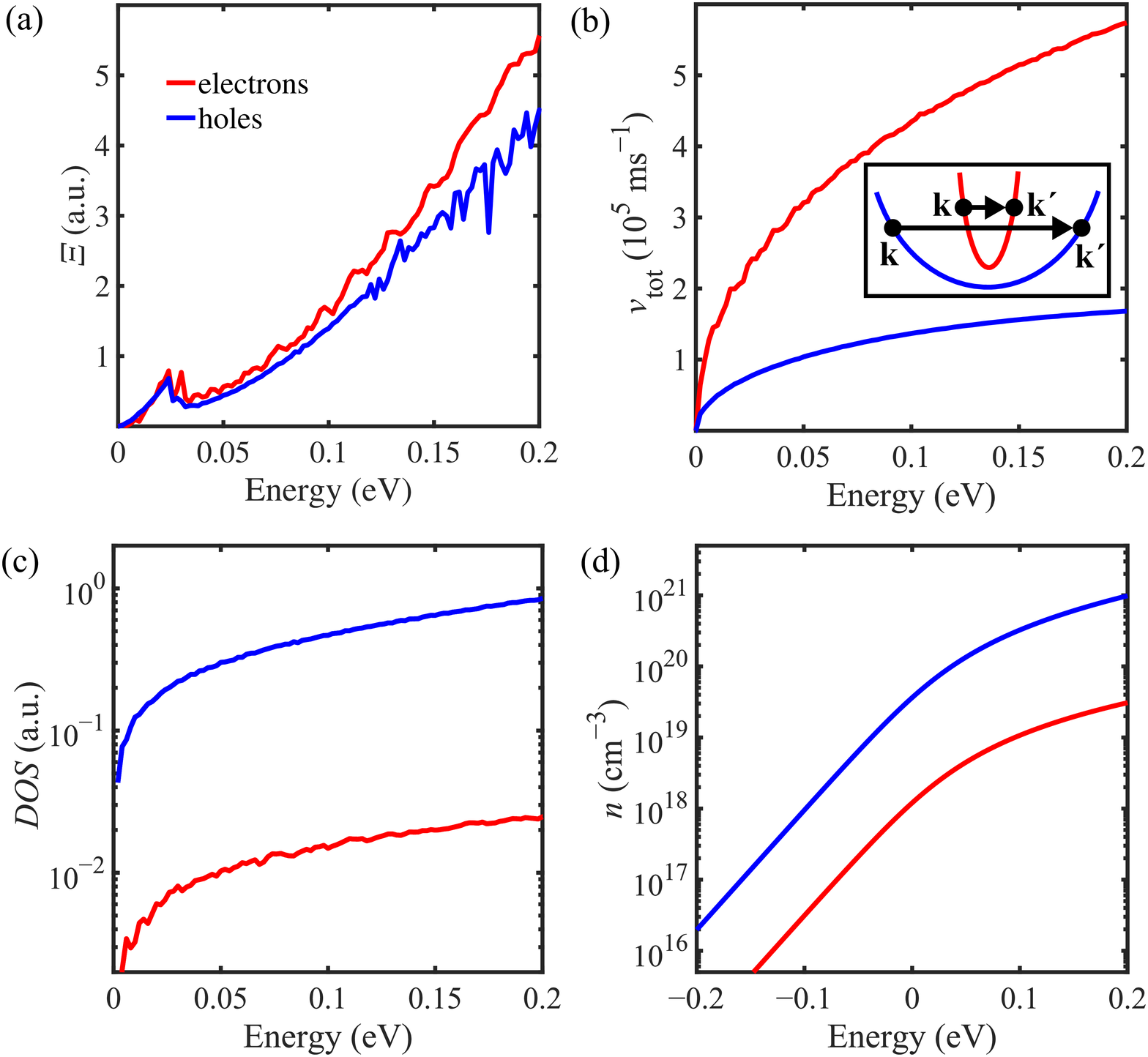}
\caption{Calculated (\textbf{a}) transport distribution functions, (\textbf{b}) band velocities, (\textbf{c}) density of states, and (\textbf{d}) carrier concentrations for electrons and holes in SnO$_2$. In (\textbf{a}), the transport distribution functions are averaged from $\Xi_{xx}$, $\Xi_{yy}$, and $\Xi_{zz}$. The inset of (\textbf{b}) shows the different $|{\bf k}-{\bf k'}|^2$ for the heavy band and light band.
\label{fig5}}
\end{figure*}

\subsection{Mobility Calculations}

The mobility is computed using the transport distribution function within the linearized Boltzmann transport equation as:
\begin{eqnarray}
\mu = \frac{q_0^2}{ne} \int_E \Xi _{ij} (E) (-\frac{\partial f_0}{\partial E}) dE
\end{eqnarray}
where $q_0$ is the electronic charge, $f_0$ is the equilibrium Fermi distribution, and $\Xi _{ij} (E)$ is the transport distribution function, which is expressed as a surface integral over the constant energy surfaces for each band and then summed over the bands:
\begin{eqnarray}
\Xi _{ij} (E) = \frac{s}{(2\pi)^3} \sum_{{\bf k},n} v_{i({\bf k},n)} v_{j({\bf k},n)} \tau_{i({\bf k},n)} g(E)
\end{eqnarray}
where $s$ is the spin degeneracy and $s$ = 2 is used as the two spin sub-bands are degenerate, $v_{i({\bf k},n)}$ is the $i$-component of the band velocity, $\tau_{i({\bf k},n)}$ is the overall relaxation time which is derived from the scattering rate $|S_{{\bf k},{\bf k'}}|$ between the considered state and all the possible final states \cite{Graziosi2019}, where $|S_{{\bf k},{\bf k'}}|$ is calculated combining the strength of all scattering mechanisms using Matthiessen's rule as~\cite{Matthiessen1864}:
\begin{eqnarray}
|S_{{\bf k},{\bf k'}}| = |S_{{\bf k},{\bf k'}}^{\rm ADP}| + |S_{{\bf k},{\bf k'}}^{\rm POP}| + |S_{{\bf k},{\bf k'}}^{\rm IIS}|
\end{eqnarray}

Figure \ref{fig4}a,b show the calculated mobilities for electrons and holes, respectively, considering only electron--phonon scattering (e-ph, solid lines) or both electron--phonon and ionized impurity scattering (e-ph+IIS, dotted lines). The calculated electron--phonon scattering limited mobilities at low carrier concentrations are $\mu_{\rm e}$ = 265 cm$^2$V$^{-1}$s$^{-1}$ and \mbox{$\mu_{\rm h}$ = 7.6 cm$^2$V$^{-1}$s$^{-1}$}.

Previous calculations reported these mobilities to be $\mu_{\rm e}$ = 187 cm$^2$V$^{-1}$s$^{-1}$ (lower compared to experiments) and \mbox{$\mu_{\rm h}$ = 10.8 cm$^2$V$^{-1}$s$^{-1}$} (much higher compared to experiments) \cite{Hu2019}. Considering the whole range of carrier concentrations, our predicted mobilities, including electron--phonon and ionized impurity scattering, are comparable to the mobilities from single crystals and are higher than those from thin films, as expected (see Figure \ref{fig4}) \cite{Morgan1966,Fonstad1971,Okude2008,Toyosaki2008,White2009,Yu2013,Mun2015,Fu2021}. This can be attributed to significant carrier scattering from the grain boundaries and dislocations induced by lattice mismatch between the film and substrates such as corundum Al$_2$O$_3$ and rutile TiO$_2$ \cite{Semancik1991,Rachut2014}. On the other hand, SnO$_2$ single crystals are found to have higher mobility than the epitaxial thin films \cite{Morgan1966,Fonstad1971,Galazka2014}. Using very thick self-buffer layers \cite{Fukumoto2020}, SnO$_2$ epitaxial thin films on TiO$_2$ (001) substrates are also found to have high electrons mobilities, which agrees very well with our calculated mobility with both electron--phonon and ionized impurity scattering.

In order to understand the large difference between the electron and hole mobilities, we calculate the transport distribution functions, as shown in Figure~\ref{fig5}a, which are found to be comparable for both electrons and holes, despite the fact that electrons have much higher band velocities and much lower density of states. The comparable transport distribution functions in Equation (14) for electrons and holes can be understood as follows. Compared to holes, electrons have much higher band velocities, as seen in Figure~\ref{fig5}b, due to their smaller effective mass. The calculated density of states for electrons is lower than for holes, as shown in Figure~\ref{fig5}c, which is expected from Equation (10). On the other hand, the total relaxation time is predominated by POP scattering. Because of this, from Equation (9), due to the smaller effective mass of electrons compared to holes, we can expect a much smaller exchange vector $|{\bf k}-{\bf k'}|^2$ for electrons, as shown in the inset of Figure~\ref{fig5}b. Overall, the larger velocity ($v^2$) and lower density of states for electrons in Equation (14) will be somewhat compensated by the smaller $|{\bf k}-{\bf k'}|^2$ and lower density of states in Equation (9) for POP scattering rates. This finally results in the comparable transport distribution functions for electrons and holes (note that although the deformation potentials for electrons and holes are very different, ADP is an overall weaker mechanism in this case). Thus, the much larger mobility for electrons compared to holes comes from the much lower carrier concentration, as shown in Figure~\ref{fig5}d, which is expected from the much smaller effective mass of the~electrons.


\section{Conclusions}

In conclusion, we employed the numerically solved linearized Boltzmann transport equation with first-principles calculated band structures to calculate the mobilities of electrons and holes in SnO$_2$. We consider acoustic deformation potential, polar optical phonon, and ionized impurity scattering processes. Both electron and hole mobilities are found to be predominantly limited by the polar optical phonon scattering at room temperature. The calculated effective masses of electrons and holes are directly related to the difference in mobilities observed between $n$-type and $p$-type SnO$_2$. The mobilities, as a function of the carrier concentration, show an upper limit of $\mu_{\rm e}$ = 265 cm$^2$V$^{-1}$s$^{-1}$ and $\mu_{\rm h}$ = 7.6 cm$^2$V$^{-1}$s$^{-1}$, which agrees well with previous experimental values, at least for the $n$-type~SnO$_2$.

\begin{acknowledgments}
This work has received funding from the European Research Council (ERC) under the European Union's Horizon 2020 research and innovation programme (grant agreement No 678763).
\end{acknowledgments}

\bibliography{mdpi}

\end{document}